\documentclass[showpacs,aps,prc,preprint,nofootinbib,superscriptaddress]{revtex4-1}

\usepackage{amsmath}
\usepackage{amssymb}
\usepackage{slashed}
\def\bm{\boldsymbol}
\newcommand{\bea}{\begin{eqnarray}}
\newcommand{\eea}{\end{eqnarray}}
\newcommand{\be}{\begin{eqnarray}}
\newcommand{\ee}{\end{eqnarray}}
\newcommand{\no}{\nonumber \\}

\newcommand{\del}{\partial}

\def\vp{{\bm p}}

\def\vx{{\bm x}}
\def\vy{{\bm y}}

\def\vr{{\bm r}}
\def\vs{{\bm\sigma}}
\def\la{\langle}
\def\ra{\rangle}

\newcommand{\sixjsymbol}[6]{\left\{\begin{tabular}{ccc} {$#1$}&{$#2$}&{$#3$}\\
                             {$#4$}&{$#5$}&{$#6$} \end{tabular}\right\}}

\begin{document}


\title { Time Reversal Invariance Violation in Neutron Deuteron  Scattering}


\author{Young-Ho Song}
\email[]{song25@mailbox.sc.edu}
\affiliation{Department of Physics and Astronomy, University of South Carolina, Columbia, SC, 29208}

\author{Rimantas Lazauskas}
\email[]{rimantas.lazauskas@ires.in2p3.fr}
\affiliation{IPHC, IN2P3-CNRS/Universit\'e Louis Pasteur BP 28,
F-67037 Strasbourg Cedex 2, France}

\author{Vladimir Gudkov}
\email[]{gudkov@sc.edu}
\affiliation{Department of Physics and Astronomy, University of South Carolina, Columbia, SC, 29208}




\date{\today}

\begin{abstract}
Time reversal invariance violating (TRIV) effects  for low energy
elastic neutron deuteron scattering
are calculated for meson exchange and EFT-type of TRIV potentials in a Distorted Wave Born Approximation,
using realistic hadronic strong interaction wave functions, obtained by solving
three-body Faddeev equations in configuration space.
The  relation  between TRIV and parity violating observables
are discussed.
\end{abstract}


\maketitle

\section{Introduction
\label{sec:Intro}}

A search for Time Reversal Invariance Violation (TRIV) in nuclear physics has been a subject of experimental and theoretical investigation  for several decades. It has covered a large variety of nuclear reactions and nuclear decays with T-violating parameters which are sensitive to either  CP-odd and P-odd (or T- and P-violating) interactions  or T-violating P-conserving (C-odd and P-even) interactions.
There are a number of advantages of the search for TRIV in nuclear processes. The main advantage is the possibility of enhancement of T-violating observables by many orders of a magnitude  due to complex nuclear structure  (see, i.e. paper \cite{Gudkov:1991qg} and references therein). Another advantage to be mentioned is
the availability of many systems with T-violating parameters which provides assurance to have enough observations against possible ``accidental'' cancellation of T-violating effects due to unknown structural factors related to strong interactions.  Taking into account that different models of CP-violation may contribute differently to a particular T/CP-observable \footnote{For example, QCD $\theta$-term can contribute to neutron EDM, but cannot be observed in $K^0$-meson decays. On the other hand, the CP-odd phase of Cabibbo-Kobayashi-Maskawa matrix was measured in  $K^0$-meson decays, but its contribution to neutron EDM is extremely small and beyond the reach with the current experimental accuracy.}, which  may have  unknown theoretical uncertainties, TRIV nuclear processes shall
provide complementary information to electric dipole moments (EDM) measurements.

One  promising approach for a search for TRIV in nuclear reactions is a measurement of TRIV effects in transmission of polarized neutron through polarized target. These effects could be measured at new  spallation neutron facilities, such as the SNS at the Oak Ridge National Laboratory or the J-SNS at J-PARC, Japan.
It was shown that these TRIV effects can be  enhanced \cite{Bunakov:1982is} by a factor of $10^6$.   Similar enhancement factors have been observed for parity violating effects in neutron scattering.  In contrast to the parity violating (PV) case, the enhancement   of TRIV effects lead not only to the opportunity to observe T violation, but also to select models of CP-violation based on the values of observed parameters.
However,  existing estimates of CP-violating effects in nuclear reactions have at least one order of  magnitude of accuracy, or even worse.
In this relation, it is interesting to compare
the calculation of TRIV effects in complex nuclei with
the  calculations of these effects in simplest few body systems,   which could be useful for clarification of influence of nuclear structure on values of TRIV effects.
Therefore, as a first step to  many body nuclear effects, we study TRIV and parity violating effects in one of the simplest available
 nuclear process, namely elastic neutron-deuteron scattering.

We treat TRIV nucleon-nucleon  interactions as a  perturbation,
while non-perturbed  three-body  wave functions are obtained by solving
 Faddeev equations  for realistic strong interaction Hamiltonian, based on
 AV18+UIX interaction model.
For description of TRIV potentials, we use both meson exchange model
and effective field theory (EFT) approach.

\section{Observables}
\label{sec:observables}
 We consider TRIV and PV effects related to $\vs_n\cdot({\vp}\times{\bm I})$ correlation, where  $\vs_n$ is the neutron spin, ${\bm I}$ is the target spin,
and $\vp$ is the neutron momentum, which can be observed in the transmission of polarized neutrons through a target with polarized nuclei.
 This correlation leads to the
difference \cite{Stodolsky:1982tp} between the total neutron cross sections  for $\vs_n$
parallel and anti-parallel to ${\vp}\times{\bm I}$, which is
\bea
\Delta\sigma_{\slashed{T}\slashed{P}}=\frac{4\pi}{p}{\rm Im}(f_{+}-f_{-}),
\eea
and neutron spin rotation angle \cite{Kabir:1982tp} $\phi$  around the axis
${\vp}\times{\bm I}$
\bea
\frac{d\phi_{\slashed{T}\slashed{P}}}{dz}=-\frac{2\pi N}{p}{\rm Re}(f_{+}-f_{-}).
\eea
Here, $f_{+,-}$ are the zero angle scattering amplitudes for neutrons polarized
parallel and anti-parallel to the ${\vp}\times{\bm I}$ axis, respectively,
 $z$ is the target length, and $N$ is a number of target nuclei per
unit volume.
It should be noted that these two parameters cannot be simulated by final state interactions (see, for example \cite{Gudkov:1991qg} and references therein), therefore, their measurements are an unambiguous  test of violation of time reversal invariance similar to the case of neutron electric dipole moment.

The scattering amplitudes can be represented in terms of matrix $\hat{R}$ which is related to
scattering matrix $\hat{S}$ as $\hat{R}=\hat{1}-\hat{S}$.
We define matrix element
$R^{J}_{l^\prime {\cal S}^\prime  ,l {\cal S}}=\la l^\prime {\cal S}^\prime|R^{J}|l {\cal S}\ra$,
where unprimed and  primed parameters correspond to initial and final states,
$l$ is an orbital angular momentum
between neutron and deuteron,  ${\cal S}$ is a sum of
neutron spin and deuteron total angular momentum, and $J$ is the total angular momentum of the neutron-deuteron system.
For low energy neutron scattering, one can consider only $s$- and $p$ -wave contributions, which  leads to the following expressions for the TRIV parameters
\bea
\label{eq:nsrTVPV}
\frac{1}{N}\frac{d\phi_{\slashed{T}\slashed{P}}}{dz}&=&-\frac{\pi}{2 p^2}\mbox{Re}
  \left[ \sqrt{2}R^{\frac{1}{2}}_{0\frac{1}{2},1\frac{3}{2}}
        -\sqrt{2}R^{\frac{1}{2}}_{1\frac{3}{2},0\frac{1}{2}}
        +2R^{\frac{3}{2}}_{0\frac{3}{2},1\frac{1}{2}}
        -2R^{\frac{3}{2}}_{1\frac{1}{2},0\frac{3}{2}}
\right],
\eea

\bea
\label{eq:dsTVPV}
\Delta\sigma_{\slashed{T}\slashed{P}}&=& \frac{\pi}{p^2}\mbox{Im}
  \left[ \sqrt{2}R^{\frac{1}{2}}_{0\frac{1}{2},1\frac{3}{2}}
        -\sqrt{2}R^{\frac{1}{2}}_{1\frac{3}{2},0\frac{1}{2}}
        +2R^{\frac{3}{2}}_{0\frac{3}{2},1\frac{1}{2}}
        -2R^{\frac{3}{2}}_{1\frac{1}{2},0\frac{3}{2}}
\right].
\eea
The symmetry violating $\hat{R}$ -matrix elements can be calculated with a high level of accuracy in Distorted Wave Born Approximation (DWBA)  as
\begin{equation}
R^J_{l' {\cal S}',l{\cal S}}
\simeq 4 i^{-l'+l+1} \mu p \;
{}^{(-)} \la \Psi,(l'{\cal S}')J J^z|V_{\slashed{T}\slashed{P}}|\Psi,(l{\cal S})J J^z\ra^{(+)},
\end{equation}
where  $\mu$ is a neutron-deuteron
reduced mass, $V_{\slashed{T}\slashed{P}}$ is TRIV nucleon-nucleon potential, and  $|\Psi,(l'{\cal S}')J J^z\rangle^{(\pm)}$
are  solutions of  3-body Faddeev equations in configuration space
for strong interaction Hamiltonian satisfying
outgoing (incoming) boundary condition.
The factor $i^{-l'+l}$ in this expression
is introduced to match the $R$-matrix definition
in the modified spherical harmonics convention \cite{Varshalovich}
with the  wave functions
in spherical harmonics convention used for wave-functions calculations.
The matrix elements of TRIV potential
in spherical harmonics convention are symmetric
and $R$-matrix in modified spherical harmonics convention
is antisymmetric under the exchange between initial
and final states.

For calculations of wave-functions, we used jj-coupling scheme instead of $l {\cal S}$ coupling scheme.
We can relate $R$-matrix elements in $l {\cal S}$ coupling scheme to  jj-coupling scheme using unitary transformation (see, for example \cite{Song:2010sz})
\bea
|[l_y\otimes(s_k \otimes j_x)_{\cal S}]_{J J_z}\rangle
&=&\sum_{j_y}|[j_x\otimes (l_y \otimes s_k)_{j_y}]_{J J_z}\rangle\no
& & \times (-1)^{j_x+j_y-J}(-1)^{l_y+s_k+j_x+J}
 [(2 j_y+1)(2 {\cal S}+1)]^{\frac{1}{2}}
 \sixjsymbol{l_y}{s_k}{j_y}{j_x}{J}{{\cal S}}.
\eea
Then,
\bea
R^{\frac{1}{2}}_{1\frac{3}{2},0\frac{1}{2}}
=\frac{2\sqrt{2}}{3}{\cal R}^{\frac{1}{2}}_{1\frac{1}{2},0\frac{1}{2}}-\frac{1}{3}{\cal R}^{\frac{1}{2}}_{1\frac{3}{2},0\frac{1}{2}},
\quad
R^{\frac{3}{2}}_{1\frac{1}{2},0\frac{3}{2}}
=-\frac{2}{3}{\cal R}^{\frac{3}{2}}_{1\frac{1}{2},0\frac{1}{2}}
   -\frac{\sqrt{5}}{3}{\cal R}^{\frac{3}{2}}_{1\frac{3}{2},0\frac{1}{2}}
\eea
where, ${\cal R}^J_{l' j',l j}$ is a R-matrix in
$jj$-basis.

\section{Time reversal violating  potentials
\label{sec:TVPVpot}}

The most general form of time reversal violating
and parity violating part of nucleon-nucleon Hamiltonian
in first order of relative nucleon momentum
can be written as the sum of momentum independent  and momentum dependent parts,
$H^{\slashed{T}\slashed{P}}=H^{\slashed{T}\slashed{P}}_{stat}+H^{\slashed{T}\slashed{P}}_{non-static}$
\cite{Pherzeg66},
\bea
\label{eq:static:pot}
H^{\slashed{T}\slashed{P}}_{stat}&=&g_1(r)\vs_{-}\cdot\hat{r}
                 +g_2(r)\tau_1\cdot\tau_2 \vs_{-}\cdot\hat{r}
                 +g_3(r)T_{12}^z\vs_{-}\cdot\hat{r}\no & &
                 +g_4(r)\tau_{+}\vs_{-}\cdot\hat{r}
                 +g_5(r)\tau_{-}\vs_{+}\cdot\hat{r}
\eea
\bea
\label{eq:nonstatic:pot}
H^{\slashed{T}\slashed{P}}_{non-static}
 &=&\left( g_6(r)+g_7(r)\tau_1\cdot\tau_2+g_8(r)T_{12}^z+g_9(r)\tau_{+} \right)\vs_\times\cdot\frac{\bar{\vp}}{m_N}
\no
 & &+\left(g_{10}(r)+g_{11}(r)\tau_1\cdot\tau_2+g_{12}(r)T_{12}^z+g_{13}(r)\tau_{+}\right)
 \no& &\times \left(\hat{r}\cdot\vs_\times\hat{r}
     \cdot \frac{\bar\vp}{m_N}
           -\frac{1}{3}\vs_\times\cdot \frac{\bar\vp}{m_N}\right)
  \no
 & &+g_{14}(r)\tau_{-}\Big(
    \hat{r}\cdot\vs_1\hat{r}\cdot(\vs_2\times \frac{\bar\vp}{m_N})
  +\hat{r}\cdot\vs_2\hat{r}\cdot(\vs_1\times \frac{\bar\vp}{m_N})
 \Big)\no
 & &+g_{15}(r)(\tau_1\times\tau_2)^z\vs_{+}\cdot \frac{\bar\vp}{m_N}
 \no & &+g_{16}(r)(\tau_1\times\tau_2)^z
    \left(\hat{r}\cdot\vs_{+}\hat{r}\cdot \frac{\bar\vp}{m_N}
         -\frac{1}{3}\vs_{+}\cdot \frac{\bar\vp}{m_N}
    \right),
\eea
where exact form of $g_i(r)$ depends on the details of particular theory.
Here, we consider three different approaches for description of TRIV interactions:
meson exchange model, pionless EFT, and pionful EFT.

TRIV meson exchange potential in general
  involves  exchanges of
pions ($J^P=0^-$, $m_\pi=140$ MeV), $\eta$-mesons($J^P=0^-$, $m_\eta=550$ MeV), and
$\rho$- and $\omega$-mesons ($J^P=1^-$, $m_{\rho,\omega}=770,780$ MeV).
 To derive this potential, we use strong ${\cal L}^{st}$ and TRIV ${\cal L}_{\slashed{T}\slashed{P}}$ Lagrangians, which can be written as \cite{Herczeg:1987gp,Liu:2004tq}
\bea
{\cal L}^{st}&=&g_{\pi}\bar{N} i\gamma_5\tau^a \pi^a N
               +g_{\eta }\bar{N}i\gamma_5\eta N
             \no & &  -g_{\rho }\bar{N}\left(\gamma^\mu-i\frac{\chi_V}{2 m_N}\sigma^{\mu\nu} q_\nu\right)\tau^a \rho^a_\mu N
             \no & &  -g_{\omega }\bar{N}\left(\gamma^\mu-i\frac{\chi_S}{2m_N}\sigma^{\mu\nu} q_\nu\right)\omega_\mu N,
\eea
 \bea
{\cal L}_{\slashed{T}\slashed{P}}
&=&\bar{N}[\bar{g}_\pi^{(0)} \tau^a \pi^a+\bar{g}_\pi^{(1)}\pi^0
           +\bar{g}_\pi^{(2)}(3\tau^z\pi^0-\tau^a\pi^a)]N\no
& &+\bar{N}[\bar{g}^{(0)}_\eta\eta+\bar{g}^{(1)}_\eta \tau^z \eta] N\no
& &+\bar{N}\frac{1}{2m_N}[\bar{g}_\rho^{(0)}\tau^a \rho_\mu^a
                         +\bar{g}^{(1)}_\rho \rho^0_\mu
                         +\bar{g}^{(2)}(3\tau^z\rho_\mu^0-\tau^a\rho^a_\mu )]
                         \sigma^{\mu\nu}q_\nu\gamma_5 N \no
& &+\bar{N}\frac{1}{2m_N}[\bar{g}^{(0)}_\omega\omega_\mu
                         +\bar{g}^{(1)}_\omega \tau^z \omega_\mu]
                         \sigma^{\mu\nu}q_\nu\gamma_5 N,
\eea
where $q_\nu=p_\nu-p'_\nu$,
$\chi_V$ and $\chi_S$ are iso-vector and
scalar magnetic moments of a nucleon ($\chi_V=3.70$ and $\chi_S=-0.12$), and $\bar{g}^{(i)}_\alpha$ are TRIV meson-nucleon coupling constants. Further, we use the following values for strong couplings constants: $g_{\pi }=13.07$, $\quad g_{\eta }=2.24$, $\quad g_{\rho }=2.75$, $\quad g_{\omega }=8.25$.

Meson exchange models from these Lagrangians lead to TRIV potential
\bea
V_{\slashed{T}\slashed{P}}
&=&
\left[-\frac{\bar{g}^{(0)}_\eta g_\eta}{2 m_N}
               \frac{m_\eta^2}{4\pi} Y_{1}(x_\eta)
               +\frac{\bar{g}^{(0)}_\omega g_\omega}{2 m_N}
               \frac{m_\omega^2}{4\pi}Y_{1}(x_\omega)\right]
               \vs_{-}\cdot\hat{r}
               \no
& &+\left[-\frac{\bar{g}^{(0)}_\pi g_\pi}{2 m_N}
              \frac{m_\pi^2}{4\pi} Y_{1}(x_\pi)
              +\frac{\bar{g}^{(0)}_\rho g_\rho}{2 m_N}
              \frac{m_\rho^2}{4\pi}Y_{1}(x_\rho)\right]
              \tau_1\cdot\tau_2\vs_{-}\cdot\hat{r}
              \no
& &+\left[-\frac{\bar{g}^{(2)}_\pi g_\pi}{2 m_N}
              \frac{m_\pi^2}{4\pi}Y_{1}(x_\pi)
              +\frac{\bar{g}^{(2)}_\rho g_\rho}{2 m_N}
              \frac{m_\rho^2}{4\pi}Y_{1}(x_\rho)\right]
              T_{12}^z\vs_{-}\cdot\hat{r}
              \no
& &+\left[-\frac{\bar{g}^{(1)}_\pi g_\pi}{4 m_N}
               \frac{m_\pi^2}{4\pi} Y_{1}(x_\pi)
              +\frac{\bar{g}^{(1)}_\eta g_\eta}{4 m_N}
               \frac{m_\eta^2}{4\pi} Y_{1}(x_\eta)
              +\frac{\bar{g}^{(1)}_\rho g_\rho}{4 m_N}
                \frac{m_\rho^2}{4\pi}Y_{1}(x_\rho)
              +\frac{\bar{g}^{(1)}_\omega g_\omega}{2 m_N}
               \frac{m_\omega^2}{4\pi} Y_{1}(x_\omega)\right]
               \tau_{+}\vs_{-}\cdot\hat{r}
              \no
& &+\left[-\frac{\bar{g}^{(1)}_\pi g_\pi}{4 m_N}
              \frac{m_\pi^2}{4\pi} Y_{1}(x_\pi)
              -\frac{\bar{g}^{(1)}_\eta g_\eta}{4 m_N}
              \frac{m_\eta^2}{4\pi} Y_{1}(x_\eta)
              -\frac{\bar{g}^{(1)}_\rho g_\rho}{4 m_N}
              \frac{m_\rho^2}{4\pi}Y_{1}(x_\rho)
              +\frac{\bar{g}^{(1)}_\omega g_\omega}{2 m_N}
              \frac{m_\omega^2}{4\pi}Y_{1}(x_\omega)\right]
              \tau_{-}\vs_{+}\cdot\hat{r},\no
\eea
where $T_{12}^z=3\tau_1^z\tau_2^z-\tau_1\cdot\tau_2$,
$Y_1(x)=(1+\frac{1}{x})\frac{e^-x}{x}$,  $x_a=m_a r$.

Comparing  eq. (\ref{eq:static:pot}) with this potential, one can see that $g_i(r)$ functions in
meson exchange model are defined as
\bea
\label{eq:gi:ME}
g_1^{ME}(r)&=&-\frac{\bar{g}^{(0)}_\eta g_\eta}{2 m_N}
               \frac{m_\eta^2}{4\pi} Y_{1}(x_\eta)
               +\frac{\bar{g}^{(0)}_\omega g_\omega}{2 m_N}
               \frac{m_\omega^2}{4\pi}Y_{1}(x_\omega)\no
g_2^{ME}(r)&=&-\frac{\bar{g}^{(0)}_\pi g_\pi}{2 m_N}
              \frac{m_\pi^2}{4\pi} Y_{1}(x_\pi)
              +\frac{\bar{g}^{(0)}_\rho g_\rho}{2 m_N}
              \frac{m_\rho^2}{4\pi}Y_{1}(x_\rho)
              \no
g_3^{ME}(r)&=&-\frac{\bar{g}^{(2)}_\pi g_\pi}{2 m_N}
              \frac{m_\pi^2}{4\pi}Y_{1}(x_\pi)
              +\frac{\bar{g}^{(2)}_\rho g_\rho}{2 m_N}
              \frac{m_\rho^2}{4\pi}Y_{1}(x_\rho)
              \no
g_4^{ME}(r)&=&-\frac{\bar{g}^{(1)}_\pi g_\pi}{4 m_N}
               \frac{m_\pi^2}{4\pi} Y_{1}(x_\pi)
              +\frac{\bar{g}^{(1)}_\eta g_\eta}{4 m_N}
               \frac{m_\eta^2}{4\pi} Y_{1}(x_\eta)
              +\frac{\bar{g}^{(1)}_\rho g_\rho}{4 m_N}
                \frac{m_\rho^2}{4\pi}Y_{1}(x_\rho)
              +\frac{\bar{g}^{(1)}_\omega g_\omega}{2 m_N}
               \frac{m_\omega^2}{4\pi} Y_{1}(x_\omega)
              \no
g_5^{ME}(r)&=&-\frac{\bar{g}^{(1)}_\pi g_\pi}{4 m_N}
              \frac{m_\pi^2}{4\pi} Y_{1}(x_\pi)
              -\frac{\bar{g}^{(1)}_\eta g_\eta}{4 m_N}
              \frac{m_\eta^2}{4\pi} Y_{1}(x_\eta)
              -\frac{\bar{g}^{(1)}_\rho g_\rho}{4 m_N}
              \frac{m_\rho^2}{4\pi}Y_{1}(x_\rho)
              +\frac{\bar{g}^{(1)}_\omega g_\omega}{2 m_N}
              \frac{m_\omega^2}{4\pi}Y_{1}(x_\omega),\no
\eea

For TRIV potentials in pionless EFT potential, these functions  are
\bea
\label{eq:gi:pionless}
g_1^{\not\pi}(r)&=& \frac{c_1^{\not\pi}}{2 m_N} \frac{d}{dr}\delta^{(3)}({\vr})\to
    -\frac{c_1^{\not\pi} \mu^2}{2 m_N} \frac{\mu^2}{4\pi}Y_1(\mu r)\no
g_2^{\not\pi}(r)&=& \frac{c_2^{\not\pi}}{2 m_N} \frac{d}{dr}\delta^{(3)}({\vr})\to -\frac{c_2^{\not\pi} \mu^2}{2 m_N} \frac{\mu^2}{4\pi}Y_1(\mu r)\no
g_3^{\not\pi}(r)&=& \frac{c_3^{\not\pi}}{2 m_N} \frac{d}{dr}\delta^{(3)}({\vr})\to -\frac{c_3^{\not\pi} \mu^2}{2 m_N} \frac{\mu^2}{4\pi}Y_1(\mu r)\no
g_4^{\not\pi}(r)&=& \frac{c_4^{\not\pi}}{2 m_N} \frac{d}{dr}\delta^{(3)}({\vr})\to -\frac{c_4^{\not\pi} \mu^2}{2 m_N} \frac{\mu^2}{4\pi}Y_1(\mu r)\no
g_5^{\not\pi}(r)&=& \frac{c_5^{\not\pi}}{2 m_N} \frac{d}{dr}\delta^{(3)}({\vr})\to -\frac{c_5^{\not\pi} \mu^2}{2 m_N} \frac{\mu^2}{4\pi}Y_1(\mu r),
\eea
where
 low energy constants (LECs) $c_i^{\not\pi}$ of pionless EFT have the dimension $[fm^2]$. In our calculations with this potential,
we use  Yukawa function ($\frac{\mu^3}{4\pi}Y_0(\mu r)$, where $Y_0(x)=\frac{e^{-x}}{x}$) with regularization scale $\mu = m_\pi$,  instead of singular $\delta^{(3)}(r)$ in paper \cite{Liu:2004tq}.

The pionful EFT acquire long range terms due to the one pion exchange in addition to
the short range term expressions equivalent to ones provided by the pionless EFT. Then, ignoring  two pion exchange
contributions at the middle range and  higher order corrections, one can write $g_i(r)$ functions for the pionful EFT as
\bea
\label{eq:gi:pionful}
g_1^{\pi}(r)&=& -\frac{c_1^{\pi} \mu^2}{2 m_N} \frac{\mu^2}{4\pi}Y_1(\mu r)\no
g_2^{\pi}(r)&=& -\frac{c_2^{\pi} \mu^2}{2 m_N} \frac{\mu^2}{4\pi}Y_1(\mu r)-\frac{\bar{g}^{(0)}_\pi g_\pi}{2 m_N}
              \frac{m_\pi^2}{4\pi} Y_{1}(x_\pi)\no
g_3^{\pi}(r)&=& -\frac{c_3^{\pi} \mu^2}{2 m_N} \frac{\mu^2}{4\pi}Y_1(\mu r)-\frac{\bar{g}^{(2)}_\pi g_\pi}{2 m_N}
              \frac{m_\pi^2}{4\pi}Y_{1}(x_\pi)\no
g_4^{\pi}(r)&=& -\frac{c_4^{\pi} \mu^2}{2 m_N} \frac{\mu^2}{4\pi}Y_1(\mu r)-\frac{\bar{g}^{(1)}_\pi g_\pi}{4 m_N}
               \frac{m_\pi^2}{4\pi} Y_{1}(x_\pi)\no
g_5^{\pi}(r)&=& -\frac{c_5^{\pi} \mu^2}{2 m_N} \frac{\mu^2}{4\pi}Y_1(\mu r)
-\frac{\bar{g}^{(1)}_\pi g_\pi}{4 m_N}
              \frac{m_\pi^2}{4\pi} Y_{1}(x_\pi).
\eea
For this potential, the cutoff scale $\mu$ is larger than pion mass, because pion  is a degree of freedom of the theory. Therefore, in general
magnitudes of LECs and their scaling behavior, as a function of a cutoff parameter
$c_i^{\pi}(\mu)$, are different from $c_i^{\not\pi}(\mu)$ ones.

One can see that  all these three potentials which come from different approaches have exactly the same operator structure. The only difference between them  is related in different scalar  function multiplied by each operator, which, in turn, defer only by different  scales of characteristic masses: $m_\pi$, $m_\eta$, $m_\rho$, and $m_\omega$.
Therefore, to unify notations,  it is convenient to define new  constants $C_n^a$(of dimension of $[fm]$)
and scalar function $f_n^a(r)=\frac{\mu^2}{4\pi}Y_1(\mu r)$ (of dimension of $[fm^{-2}]$) as
\bea
g_n(r)\equiv \sum_{a} C_n^a f_n^a(r),
\eea
where the form of $C_n^a$ and $f_n^a(r)$ can be read from eq.
(\ref{eq:gi:ME}), (\ref{eq:gi:pionless}) and (\ref{eq:gi:pionful}).

Since non-static TRIV potentials, with $g_{n>5}$,
do not appear  either in meson exchange model or in the lowest order EFTs,
they can be considered as a higher order correction to the lowest order EFT
or related to heavy meson  contributions in the meson exchange model.
Nevertheless, for a completeness of consideration, we estimate the contributions of these operators
 using $f_n^a(r)$ functions with proper mass scales.

\section{Calculation of TRIV amplitudes
\label{sec:matrix}}

The non-perturbed (parity conserving) 3-body wave functions for neutron-deuteron scattering
are obtained by solving Faddeev equations (also often called Kowalski-Noyes
equations) in configuration space~\cite{Faddeev:1960su,Lazauskas:2004hq}.
The wave function in Faddeev formalism is a sum of three Faddeev
components,
\bea
\Psi({\bm x},{\bm y})=\psi_1({\bm x}_1,{\bm y}_1)
                      +\psi_2({\bm x}_2,{\bm y}_2)
                      +\psi_3({\bm x}_3,{\bm y}_3).
\eea
In a particular case of three identical particles (this becomes formally
true for three-nucleon system in the isospin formalism),
three Faddeev equations (components)
become formally identical. By accommodating the three-nucleon
force, which under nucleon permutation might be expressed as a symmetric sum of three terms:
$V_{ijk}=V_{ij}^{k}+V_{jk}^{i}+V_{ki}^{j},$
Faddeev equations read:
\begin{equation}
\left( E-H_{0}-V_{ij}\right) \psi_{k}=V_{ij}(\psi_{i}+\psi_{j})+\frac{1}{2}(V_{jk}^{i}+V_{ki}^{j})\Psi
\label{EQ_FE}
\end{equation}
where $(ijk)$ are particle indices, $H_{0}$ is kinetic energy operator,
$V_{ij}$ is two body force between particles $i$, and $j$, and
$\psi_{k}=\psi_{ij,k}$ is
Faddeev component.

Using relative Jacobi coordinates
$\vx_{k}=(\vr_{j}-\vr_{i})\smallskip $
and
$\vy_{k}=
\frac{2}{\sqrt{3}}(\vr_{k}-\frac{\vr_{i}+\vr_{j}}{2})$,
one can expand these  Faddeev components in bipolar harmonic basis:
\begin{equation}
\psi _{k}=\sum\limits_{\alpha }\frac{F_{\alpha }(x_{k},y_{k})}{x_{k}y_{k}}%
\left\vert \left( l_{x}\left( s_{i}s_{j}\right) _{s_{x}}\right)
_{j_{x}}\left( l_{y}s_{k}\right) _{j_{y}}\right\rangle _{JM}\otimes
\left\vert \left( t_{i}t_{j}\right) _{t_{x}}t_{k}\right\rangle _{TT_{z}},
\label{EQ_FA_exp}
\end{equation}%
where index $\alpha $ represents all allowed combinations of the
quantum numbers presented in the brackets: $l_{x}$ and $l_{y}$ are the
partial angular momenta associated with respective Jacobi coordinates, $%
s_{i} $ and $t_{i}$ are the spins and isospins of the individual particles. Functions
$F_{\alpha }(x_{k},y_{k})$ are called partial Faddeev amplitudes.
It should be noted that the total angular momentum $J$ as well as its
projection $M$ are conserved, but the total isospin $T$ of the system is not conserved due to the presence of charge dependent terms in nuclear
interactions.

 Boundary conditions for Eq.~(\ref{EQ_FE}) can be written in the
Dirichlet form. Thus, Faddeev amplitudes satisfy the regularity conditions:
\begin{equation}
F_{\alpha }(0,y_{k})=F_{\alpha }(x_{k},0)=0.  \label{BC_xyz_0}
\end{equation}%
For neutron-deuteron scattering
with energies below the break-up threshold, Faddeev components
vanish  for $\mathbf{x}_{k}\rightarrow\infty$.
If $\mathbf{y}_{k}\rightarrow\infty $, then interactions between the particle $k$ and the cluster $ij$
are negligible, and Faddeev components $\psi_{i}$ and
$\psi_{j}$ vanish. Then, for
the component $\psi_{k}$, which describes
the plane wave of the particle $k$ with respect
to the bound particle pair $ij$,
\begin{eqnarray}
\lim_{y_k\to \infty}
\psi_{k}(\mathbf{x}_{k},\mathbf{y}_{k} )_{l_{n}j_{n}}&=&
\frac{1}{\sqrt{3}}\sum\limits_{j_{n}^{\prime }l_{n}^{\prime }}
\left| \left\{\phi_{d}(\mathbf{x}_{k})\right\}_{j_{d}}\otimes \left\{ Y_{l_{n}^{\prime }}(\mathbf{\hat{y}}_{k})\otimes s_{k}\right\}
_{j^{\prime}_{n}}\right\rangle_{JM}
\otimes \left\vert \left( t_{i}t_{j}\right)_{t_{d}}t_{k}\right\rangle_{\frac12,-\frac12}
\notag \\
&&
\times \frac{i}{2}\left[\delta_{l_{n}^{\prime}j_{n}^{%
\prime},l_{n}j_{n}}h_{l^\prime_{n}}^{-}(pr_{nd})-S_{l_{n}^{\prime}j_{n}^{%
\prime},l_{n}j_{n}}h_{l^\prime_{n}}^{+}(pr_{nd})\right],  \label{eq_as_beh}
\end{eqnarray}%
where deuteron, being formed from nucleons $i$ and $j$, has quantum numbers
$s_{d}=1$,  $j_{d}=1$, and $t_{d}=0$, and its wave function
$\phi_{d}(\mathbf{x}_{k})$ is normalized to unity. Here,
 $r_{nd}=(\sqrt{3}/2)y_{k}$ is the relative distance between
neutron and deuteron target, and $h_{l_{n}}^{\pm }$ are
the spherical Hankel
functions. The expression~(\ref{eq_as_beh}) is normalized to satisfy a condition of  unit flux for
$nd$ scattering
wave function.

For the cases where Urbana type three-nucleon interaction (TNI) is included,
we modify the Faddeev equation (\ref{EQ_FE}) into
\begin{equation}
\left( E-H_{0}-V_{ij}\right) \psi_{k}=V_{ij}(\psi_{i}+\psi_{j})+\frac{1}{2}(V_{jk}^{i}+V_{ki}^{j})\Psi
\end{equation}%
by noting that the TNI among particles $ijk$ can be written as the sum of three terms:
$V_{ijk}=V_{ij}^{k}+V_{jk}^{i}+V_{ki}^{j}$.

Using decomposition of momentum ${\bar\vp}$ which
acts only on the nuclear wave function,
\bea
\bar{\vp}=\frac{i\overleftarrow{\nabla}_x-i\overrightarrow{\nabla}_x}{2}
         =\frac{i\hat{x}}{2}\left(\overleftarrow{\frac{\del}{\del x}}
                                 -\overrightarrow{\frac{\del}{\del x}}
         \right) +\frac{i}{2}\frac{1}{x}\left(
        \overleftarrow{\nabla}_\Omega-i\overrightarrow{\nabla}_\Omega
         \right),
\eea
we can represent general matrix elements of local two-body
parity violating potential operators as
\bea
{}^{(-)}\la \Psi_f|O|\Psi_i\ra^{(+)}
=
(\frac{\sqrt{3}}{2})^3\sum_{\alpha\beta}\left[\int dx x^2 dy y^2
 \left(\frac{\widetilde{F}^{(+)}_{f,\alpha}(x,y)}{xy}\right)
 \hat{X}(x)
 \left(\frac{\widetilde{F}^{(+)}_{i,\beta}(x,y)}{xy}\right)
 \right] \la \alpha|\hat{O}(\hat{x})|\beta\ra,\no
\eea
where
$(\pm)$ means outgoing and incoming boundary conditions
and $\hat{X}(x)$ is a scalar function or
derivative acting on wave function with respect to $x$.
(Note that we have used the fact that $(\widetilde{F}^{(-)})^*=\widetilde{F}^{(+)}$.)
The partial amplitudes $\widetilde{F}_{i(f),\alpha}(x,y)$ represent
the total systems wave function in
one selected basis set among three possible angular momentum coupling
sequences for three particle angular momenta:
\begin{equation}
\Psi_{i(f)}(x,y)=\sum\limits_{\alpha }\frac{\widetilde{F}_{i(f),{\alpha }}(x,y)}{xy}%
\left\vert \left( l_{x}\left( s_{i}s_{j}\right) _{s_{x}}\right)
_{j_{x}}\left( l_{y}s_{k}\right) _{j_{y}}\right\rangle _{JM}\otimes
\left\vert \left( t_{i}t_{j}\right) _{t_{x}}t_{k}\right\rangle _{TT_{z}}.
\label{EQ_FW_exp}
\end{equation}

The ``angular'' part of the matrix element is
\bea
\la \alpha|\hat {O}(\hat{x})|\beta\ra
\equiv \int d\hat{x}\int d\hat{y}
{\cal Y}^\dagger_\alpha(\hat{x},\hat{y})\hat{O}(\hat{x})
{\cal Y}_\beta(\hat{x},\hat{y}),
\eea
where ${\cal Y}_\alpha(\hat{x},\hat{y})$
is a tensor bipolar spherical harmonic
with a quantum number $\alpha$.
One can see that operators for ``angular" matrix elements  have the following structure:
\bea
\hat{O}(\hat{x})=(\tau_i \odot\tau_j)(\vs_i\circledcirc\vs_j)\cdot
(\hat{x},\mbox{ or }
\overleftarrow{\nabla}_{\Omega},
\mbox{ or }
\overrightarrow{\nabla}_{\Omega}),
\eea
where $\odot,\circledcirc=\pm,\times$.
We calculated the ``angular" matrix elements by representing all
operators as a tensor product of isospin, spin, spatial operators.
For details of the calculations of matrix elements, see paper \cite{Song:2010sz}.
Similar approaches have been successfully applied for calculations of
 weak and electromagnetic processes involving
three-body and four-body hadronic systems \cite{Song:2007bj,Song:2008zf,Lazauskas:2009nw,
Park:2002yp,Pastore:2009is,Girlanda:2010vm} and for calculation of
  parity violating effects
in neutron deuteron scattering \cite{Schiavilla:2008ic, Song:2010sz}.

\section{Results and discussions
\label{sec:results:tvpv}}

Typical results
for contributions of different operators of a TRIV potential to matrix elements  are shown in table \ref{tbl:tvpv:pion:re}, where a
 mass scale was chosen to be equal to $\mu=138\; MeV$.
As it was discussed,  both pionless and pionfull EFTs in the
leading order, as well as  the meson exchange model, have only first five operators which have non-zero values. Taking into account that the characteristic mass scale $\mu$ for operator with $g_{n\geq 6}$
should be at least larger than two-pion mass
(since  two pion exchange corresponds to  higher order corrections), the actual contributions of these operators are at least one order of magnitude smaller than the value shown in Table \ref{tbl:tvpv:pion:re}.
Thus, one can neglect contributions from the suppressed $n\geq 6$ operators provided coupling constants satisfy the naturalness assumption.

\begin{table}
\caption{\label{tbl:tvpv:pion:re} A typical matrix elements
of TRIV potential, ${\rm Re}\frac{\la  (l'_y j'_y),J|V^{\slashed{T}\slashed{P}}_n|(l_y j_y),J\ra}{\tilde{C}_n p}$,
in jj-coupling scheme with $AV18+UIX$ strong potential
at zero energy limit.
Imaginary part of potential matrix element is zero at zero energy limit.
Scalar functions are chosen as $\frac{m_\pi^2}{4\pi}Y_1(m_\pi r)$
for operators $1-5$, $\frac{m_\pi^2}{4\pi}Y_0(m_\pi r)$ for operators
$6-16$. $O_{3,8,12}=0$ because of isospin selection rules. All data are in $fm^2$.
}
\begin{ruledtabular}
\begin{tabular}{ccccccccc}
  n  & $\la 1\frac{1}{2}|v^{1/2}|0\frac{1}{2}\ra/p$
     & $\la 1\frac{3}{2}|v^{1/2}|0\frac{1}{2}\ra/p$
     & $\la 1\frac{1}{2}|v^{3/2}|0\frac{1}{2}\ra/p$
     & $\la 1\frac{3}{2}|v^{3/2}|0\frac{1}{2}\ra/p$ \\
     \hline
    1   & $  0.590\times 10^{-01}  $ & $  -0.787\times 10^{-01}  $ & $   0.151\times 10^{-01}  $ & $   0.177\times 10^{-01}$\\
    2   & $  0.627\times 10^{+00}  $ & $  -0.863\times 10^{-01}  $ & $  -0.144\times 10^{+00}  $ & $  -0.167\times 10^{+00}$\\
    4   & $ -0.268\times 10^{+00}  $ & $   0.107\times 10^{+00}  $ & $   0.330\times 10^{-01}  $ & $   0.379\times 10^{-01} $\\
    5   & $  0.321\times 10^{+00}  $ & $  -0.267\times 10^{+00}  $ & $  -0.199\times 10^{+00}  $ & $  -0.691\times 10^{-01} $\\
    \hline
    6   & $  0.719\times 10^{-01}  $ & $  -0.104\times 10^{-01}  $ & $  -0.115\times 10^{-01}  $ & $  -0.141\times 10^{-01} $\\
    7   & $ -0.206\times 10^{-01}  $ & $   0.520\times 10^{-02}  $ & $   0.337\times 10^{-01}  $ & $   0.384\times 10^{-01} $\\
    9   & $ -0.650\times 10^{-01}  $ & $   0.865\times 10^{-02}  $ & $   0.238\times 10^{-03}  $ & $   0.134\times 10^{-02} $\\
   10   & $  0.106\times 10^{-01}  $ & $  -0.932\times 10^{-03}  $ & $   0.658\times 10^{-03}  $ & $   0.622\times 10^{-03} $\\
   11   & $  0.171\times 10^{-01}  $ & $  -0.548\times 10^{-03}  $ & $  -0.237\times 10^{-02}  $ & $  -0.273\times 10^{-02} $\\
   13   & $ -0.163\times 10^{-01}  $ & $   0.111\times 10^{-02}  $ & $   0.131\times 10^{-03}  $ & $   0.288\times 10^{-03} $\\
   14   & $  0.649\times 10^{-02}  $ & $  -0.628\times 10^{-02}  $ & $  -0.876\times 10^{-02}  $ & $  -0.250\times 10^{-03} $\\
   15   & $  0.338\times 10^{-01}  $ & $  -0.230\times 10^{-01}  $ & $  -0.293\times 10^{-01}  $ & $  -0.198\times 10^{-02} $\\
   16   & $  0.128\times 10^{-01}  $ & $  -0.816\times 10^{-02}  $ & $  -0.119\times 10^{-01}  $ & $  -0.335\times 10^{-03} $\\
\end{tabular}
\end{ruledtabular}
\end{table}

\begin{table}
\caption{\label{tbl:tvpv:scatt:100keV}
 Difference of scattering amplitudes,
$(f^{\slashed{T}\slashed{P}}_{+}-f^{\slashed{T}\slashed{P}}_{-})/(p C_n)$ for TRIV potential
 operators $n=1$, 2, 4, and 5 for mass scales corresponding to meson masses at $E_{cm}=100$ keV. All data are in $fm$.
}
\begin{ruledtabular}
\begin{tabular}{lrrrr}
  n  &  $\frac{\Delta f^{\pi}}{p}$  & $\frac{\Delta f^{\eta}}{p}$
     & $ \frac{\Delta f^{\rho}}{p}$ & $ \frac{\Delta f^{\omega}}{p}$  \\
\hline
    1 & $-0.615     -i0.0567 $ & $-0.317 -i0.00738$ & $-0.125    -i0.00329$ & $ -0.119    -i0.00317$   \\
    2  &$-7.58 +i 1.07$        & $ -0.761+i0.0901$  & $-0.302  +i   0.0361$ & $-0.288   +i  0.0345$    \\
    4  & $3.14 -i 0.300   $    & $0.571  -i0.0227$  & $0.225    -i0.00873$  & $0.215    -i0.00832 $     \\
    5  &$-4.99  +i0.848 $      &$ -0.262 +i0.0717$  & $-0.0934 +i 0.0273$   & $-0.0888 +i 0.0260$      \\
\end{tabular}
\end{ruledtabular}
\end{table}

The possible contributions of different mesons to TRIV amplitude
at $E_{cm}=100$ keV
are summarized in Table \ref{tbl:tvpv:scatt:100keV}. Using these data,
the observable parameters  at the neutron energy $E_{cm}=100$ keV can be re-written in terms of TRIV meson  coupling constants as
\bea
\label{eq:phiTP}
\frac{1}{N}\frac{d\phi^{\slashed{T}\slashed{P}}}{dz}&=&
(-65 \mbox{ rad}\cdot \mbox{ fm}^2)[\bar{g}_\pi^{(0)}+0.12 \bar{g}_\pi^{(1)}
+0.0072 \bar{g}_\eta^{(0)}+0.0042 \bar{g}_\eta^{(1)} \no & &
-0.0084 \bar{g}_\rho^{(0)}+0.0044 \bar{g}_\rho^{(1)}
-0.0099 \bar{g}_\omega^{(0)}+0.00064 \bar{g}_\omega^{(1)}]
\eea
and
\bea
\label{eq:PTP}
P^{\slashed{T}\slashed{P}}=\frac{\Delta\sigma^{\slashed{T}\slashed{P}}}{2\sigma_{tot}}&=&
\frac{(-0.185 \mbox{ b})}{2\sigma_{tot}}
[\bar{g}_\pi^{(0)}+0.26 \bar{g}_\pi^{(1)}
-0.0012 \bar{g}_\eta^{(0)}+0.0034 \bar{g}_\eta^{(1)} \no & &
-0.0071 \bar{g}_\rho^{(0)}+0.0035 \bar{g}_\rho^{(1)}
+0.0019 \bar{g}_\omega^{(0)}-0.00063 \bar{g}_\omega^{(1)}].
\eea

For a comparison, DDH model of PV interaction
with AV18+UIX strong potential at $E_{cm}=100$ keV gives
\bea
\label{eq:phiP}
\frac{1}{N}\frac{d\phi^{\slashed{P}}}{dz}&=&
  (55\mbox{ rad}\cdot\mbox{fm}^2)\left[
  h_\pi^1+h_\rho^0(0.11)+h_\rho^1(-0.035)
         +h_\omega^0(0.14)+h_\omega^1(-0.12)
         +h_\rho^{'1}(-0.013)
  \right]\no
  \eea
  \bea
  \label{eq:PP}
P^{\slashed{P}}=\frac{\Delta\sigma^{\slashed{P}}}{2\sigma_{tot}}&=&
  \frac{(0.395 \mbox{ b})}{2\sigma_{tot}}\left[
  h_\pi^1+h_\rho^0(0.021)+h_\rho^1(0.0027)
         +h_\omega^0(0.022)+h_\omega^1(-0.043)
         +h_\rho^{'1}(-0.012)
  \right].\no
\eea
These expressions correspond to
\bea
\frac{1}{N}\frac{d\phi^{\slashed{P}}}{dz}&=&
(59 \mbox{ rad}\cdot  \mbox{fm}^2)\left[
h_\pi^1+h_\rho^0(0.10)+h_\omega^0(0.14)\right.\no
& &\left. +h_\rho^1(-0.042)
+h_\omega^1(-0.12)+h^{'1}_\rho(0.014)\right]
\eea
for at zero energy limit, and to
\bea
P^{\slashed{P}}&=&\frac{\Delta\sigma^{\slashed{P}}}{2\sigma_{tot}}=\frac{(0.140 \mbox{ b})}{2\sigma_{tot}}
\left[
h_\pi^1+h_\rho^0(0.021)+h_\omega^0(0.022)\right.\no
& &\quad\left.
+h_\rho^1(0.002)+h_\omega^1(-0.044)
+h_\rho^{'1}(-0.012)
\right]
\eea
 at $E_{cm}=10$ keV, which were calculated with for DDH-II/AV18+UIX potentials in paper \cite{Song:2010sz}. The equations satisfy  the expected dependence of $\Delta\sigma^{\slashed{T}\slashed{P}}$ and $\Delta\sigma^{\slashed{P}}$ on neutron energy as $(E_n)^{1/2}$.
 The angle of spin rotation, being proportional to the scattering length, is not sensitive to neutron energy at low energy regime.

The results of Table \ref{tbl:tvpv:scatt:100keV} also could be considered as an illustration of the cutoff dependence of matrix elements for EFT calculations.
However, physical observables do not depend  on the cutoff
due to  the renormalization of $C^{\not\pi}_i=-\frac{c_i^{\not\pi}\mu^2}{2 m_N}$.
In pionless EFT with cutoff $\mu=m_\pi$, observables can be written
in terms of dimensional LECs, $c_i^{\not\pi}$ (in $fm^2$),
\bea
\frac{1}{N}\frac{d\phi^{\slashed{T}\slashed{P}}}{dz}
&=&(-2.45 \mbox{ rad})[c_2^{\not{\pi}}+c_1^{\not{\pi}}(0.081)+c_4^{\not{\pi}}(0.41)+c_5^{\not{\pi}}(0.66)],\no
P^{\slashed{T}\slashed{P}}=\frac{\Delta\sigma^{\slashed{T}\slashed{P}}}{2\sigma_{tot}}
&=&\frac{(-0.35 )}{\sigma_{tot}}
  [c_2^{\not{\pi}}+c_1^{\not{\pi}}(-0.053)+c_4^{\not{\pi}}(-0.28)+c_5^{\not{\pi}}(0.79)].
\eea

For the case of pionful EFT, one pion exchange contribution
is taken explicitly, and all other
cutoffs for contact terms should be larger than pion mass.
Therefore, the results in table \ref{tbl:tvpv:scatt:100keV} for pion, $\rho$, and $\omega$ masses correspond to results for different $\mu$'s. For example, choosing cutoff scale  $\mu = m_\rho$,
the expressions for TRIV observables are
\bea
\frac{1}{N}\frac{d\phi^{\slashed{T}\slashed{P}}}{dz}&=&
(-65 \mbox{ rad}\cdot \mbox{ fm}^2)[\bar{g}_\pi^{(0)}+0.12 \bar{g}_\pi^{(1)}] \no & &
+(-3.05\mbox{ rad})[c_2^\pi + c_1^\pi (0.41) + c_4^\pi ( -0.75)+c_5^\pi (0.31) ]
\eea
and
\bea
P^{\slashed{T}\slashed{P}}=\frac{\Delta\sigma^{\slashed{T}\slashed{P}}}{2\sigma_{tot}}&=&
\frac{(-0.185 \mbox{ b})}{2\sigma_{tot}}
[\bar{g}_\pi^{(0)}+0.26 \bar{g}_\pi^{(1)}]\no & &
+\frac{(-0.728)}{2\sigma_{tot}}[c_2^\pi+ c_1^\pi(-0.091)
                        + c_4^\pi (-0.24)+ c_5^\pi (0.76)].
\eea

It should be noted that all existing calculation of TRIV couplings are
based on the meson exchange model, since EFT low energy constants
for TRIV interactions are unknown.
Using meson exchange model,
one can predict  TRIV effects
for different models of CP-violation mechanism, because values of TRIV meson-nucleon coupling constants depend on models of CP-violation.

The results of the calculations show that the dominant contributions to TRIV effects come from the first five operators.
Moreover, in  meson exchange formalism, pion exchange contribution is dominant, provided that CP-odd coupling constants for all mesons have the same order of magnitude. Thus, comparing Eqs.(\ref{eq:phiTP}) and (\ref{eq:PTP}) with Eqs.(\ref{eq:phiP}) and  (\ref{eq:PP}), one can see that contributions from $\rho$ and $\omega$ mesons to TRIV effects are suppressed by about one order of magnitude in comparison to the contributions of these mesons to PV effects. This fact is especially interesting because, in the majority of models of CP violation, TRIV pion nucleon coupling constants  are much  larger than  $\rho$ and $\omega$ ones (for details see, for example \cite{Gudkov:1995tp,Gudkov:1991tp,Gudkov:1992yc,Pospelov:2001ys} and references therein.)  Assuming dominant contributions of $\pi$ -mesons and  using the conventional parameter \cite{Herczeg:1987gp,Gudkov:1990tb} $\lambda = \bar{g}_\pi / h_\pi^1$, one can describe the TRIV observable in terms of corresponding PV ones as
\bea
\label{eq:aprox}
\frac{\phi^{\slashed{T}\slashed{P}}}{\phi^{\slashed{P}}}
  &\simeq& (1.2)\left(\frac{\bar{g}^{(0)}_\pi}{h_\pi^1}
                   +(0.12)\frac{\bar{g}^{(1)}_\pi}{h_\pi^1}\right), \no
\frac{\Delta\sigma^{\slashed{T}\slashed{P}}}{\Delta\sigma^{\slashed{P}}}
  &\simeq& (-0.47)\left(\frac{\bar{g}^{(0)}_\pi}{h_\pi^1}
                   +(0.26)\frac{\bar{g}^{(1)}_\pi}{h_\pi^1}\right).
\eea
These ratios of TRIV and PV parameters do not depend on neutron energy.

It is useful to relate these estimates to the existing experimental constrains obtained from electric dipole moment (EDM) measurements, even in the case of model dependent relations.  For example, the CP-odd coupling constant $\bar{g}^{(0)}_\pi$ could be related to the value of neutron electric dipole moment (EDM) $d_n$ generated via a $\pi^-$ -loop in the chiral limit \cite{Pospelov:2005pr} as
\bea
d_n=\frac{e}{4\pi m_N}\bar{g}^{(0)}_\pi g_\pi \ln \frac{\Lambda}{m_\pi},
\eea
where $\Lambda\simeq m_\rho$. Then, using experimental limit \cite{Baker:2006ts} on  $d_n$, one can estimate $\bar{g}^{(0)}_\pi < 2.5 \cdot 10^{-10}$. The constant $\bar{g}^{(1)}_\pi$ can be bounded using constraint \cite{Romalis:2000mg} on $^{199}Hg$ atomic EDM as $\bar{g}^{(1)}_\pi<0.5\cdot 10^{-11}$ \cite{Dmitriev:2003hs}.

 Theoretical predictions for $\lambda$ can vary from $10^{-2}$ to $10^{-10}$ for different models of CP violations (see, for example, \cite{Herczeg:1987gp,Gudkov:1990tb,Gudkov:1991tp,Gudkov:1992yc,Gudkov:1995tp} and references therein). Therefore,  one can estimate a range of  possible values of TRIV observable and relate a particular mechanism of CP-violation to their values.
It should be noted that the above parametrization assumes that pion meson exchange contribution is dominant for PV effects.  Should the $\overrightarrow{n}+p \rightarrow d+\gamma$ experiment confirm the ``best value'' of the DDH pion-nucleon coupling constant $h_\pi^1$, Eqs.(\ref{eq:aprox})  can be considered as an  estimate for the value of TRIV effects in neutron-deuteron scattering. Otherwise, if $h_\pi^1$ is small, one needs to use  $h_\rho$ or  $h_\omega$ with corresponding weights, which will increase  relative values of TRIV effects.

\begin{acknowledgments}
This work was supported by the DOE grants no. DE-FG02-09ER41621.
This work was granted access to the HPC resources of IDRIS
under the allocation 2009-i2009056006
made by GENCI (Grand Equipement National de Calcul Intensif).
We thank the staff members of the IDRIS for their constant help.
\end{acknowledgments}

\bibliography{TViolation}

\begin{thebibliography}{27}%
\makeatletter
\providecommand \@ifxundefined [1]{%
 \@ifx{#1\undefined}
}%
\providecommand \@ifnum [1]{%
 \ifnum #1\expandafter \@firstoftwo
 \else \expandafter \@secondoftwo
 \fi
}%
\providecommand \@ifx [1]{%
 \ifx #1\expandafter \@firstoftwo
 \else \expandafter \@secondoftwo
 \fi
}%
\providecommand \natexlab [1]{#1}%
\providecommand \enquote  [1]{``#1''}%
\providecommand \bibnamefont  [1]{#1}%
\providecommand \bibfnamefont [1]{#1}%
\providecommand \citenamefont [1]{#1}%
\providecommand \href@noop [0]{\@secondoftwo}%
\providecommand \href [0]{\begingroup \@sanitize@url \@href}%
\providecommand \@href[1]{\@@startlink{#1}\@@href}%
\providecommand \@@href[1]{\endgroup#1\@@endlink}%
\providecommand \@sanitize@url [0]{\catcode `\\12\catcode `\$12\catcode
  `\&12\catcode `\#12\catcode `\^12\catcode `\_12\catcode `\%12\relax}%
\providecommand \@@startlink[1]{}%
\providecommand \@@endlink[0]{}%
\providecommand \url  [0]{\begingroup\@sanitize@url \@url }%
\providecommand \@url [1]{\endgroup\@href {#1}{\urlprefix }}%
\providecommand \urlprefix  [0]{URL }%
\providecommand \Eprint [0]{\href }%
\@ifxundefined \urlstyle {%
  \providecommand \doi  [0]{\begingroup \@sanitize@url \@doi}%
  \providecommand \@doi [1]{\endgroup \@@startlink {\doibase
  #1}doi:\discretionary {}{}{}#1\@@endlink }%
}{%
  \providecommand \doi  [0]{doi:\discretionary{}{}{}\begingroup
  \urlstyle{rm}\Url }%
}%
\providecommand \doibase [0]{http://dx.doi.org/}%
\providecommand \Doi [0]{\begingroup \@sanitize@url \@Doi }%
\providecommand \@Doi  [1]{\endgroup\@@startlink{\doibase#1}\@@Doi}%
\providecommand \@@Doi [1]{#1\@@endlink}%
\providecommand \selectlanguage [0]{\@gobble}%
\providecommand \bibinfo  [0]{\@secondoftwo}%
\providecommand \bibfield  [0]{\@secondoftwo}%
\providecommand \translation [1]{[#1]}%
\providecommand \BibitemOpen [0]{}%
\providecommand \bibitemStop [0]{}%
\providecommand \bibitemNoStop [0]{.\EOS\space}%
\providecommand \EOS [0]{\spacefactor3000\relax}%
\providecommand \BibitemShut  [1]{\csname bibitem#1\endcsname}%
\bibitem [{\citenamefont {Gudkov}(1992){\natexlab{a}}}]{Gudkov:1991qg}%
  \BibitemOpen
  \bibfield  {author} {\bibinfo {author} {\bibfnamefont {V.~P.}\ \bibnamefont
  {Gudkov}},\ }\href@noop {} {\bibfield  {journal} {\bibinfo  {journal} {Phys.
  Rept.},\ }\textbf {\bibinfo {volume} {212}},\ \bibinfo {pages} {77} (\bibinfo
  {year} {1992}{\natexlab{a}})}\BibitemShut {NoStop}%
\bibitem [{\citenamefont {Bunakov}\ and\ \citenamefont
  {Gudkov}(1983)}]{Bunakov:1982is}%
  \BibitemOpen
  \bibfield  {author} {\bibinfo {author} {\bibfnamefont {V.~E.}\ \bibnamefont
  {Bunakov}}\ and\ \bibinfo {author} {\bibfnamefont {V.~P.}\ \bibnamefont
  {Gudkov}},\ }\href@noop {} {\bibfield  {journal} {\bibinfo  {journal} {Nucl.
  Phys.},\ }\textbf {\bibinfo {volume} {A401}},\ \bibinfo {pages} {93}
  (\bibinfo {year} {1983})}\BibitemShut {NoStop}%
\bibitem [{\citenamefont {Stodolsky}(1982)}]{Stodolsky:1982tp}%
  \BibitemOpen
  \bibfield  {author} {\bibinfo {author} {\bibfnamefont {L.}~\bibnamefont
  {Stodolsky}},\ }\href@noop {} {\bibfield  {journal} {\bibinfo  {journal}
  {Nucl. Phys.},\ }\textbf {\bibinfo {volume} {B197}},\ \bibinfo {pages} {213}
  (\bibinfo {year} {1982})}\BibitemShut {NoStop}%
\bibitem [{\citenamefont {Kabir}(1982)}]{Kabir:1982tp}%
  \BibitemOpen
  \bibfield  {author} {\bibinfo {author} {\bibfnamefont {P.~K.}\ \bibnamefont
  {Kabir}},\ }\href@noop {} {\bibfield  {journal} {\bibinfo  {journal} {Phys.
  Rev.},\ }\textbf {\bibinfo {volume} {D25}},\ \bibinfo {pages} {2013}
  (\bibinfo {year} {1982})}\BibitemShut {NoStop}%
\bibitem [{\citenamefont {Varshalovich}\ \emph {et~al.}(1988)\citenamefont
  {Varshalovich}, \citenamefont {Moskalev},\ and\ \citenamefont
  {Khersonskii}}]{Varshalovich}%
  \BibitemOpen
  \bibfield  {author} {\bibinfo {author} {\bibfnamefont {D.~A.}\ \bibnamefont
  {Varshalovich}}, \bibinfo {author} {\bibfnamefont {A.~N.}\ \bibnamefont
  {Moskalev}}, \ and\ \bibinfo {author} {\bibfnamefont {V.~K.}\ \bibnamefont
  {Khersonskii}},\ }\href@noop {} {\emph {\bibinfo {title} {Quantum Theory of
  Angular Momentum}}}\ (\bibinfo  {publisher} {World Scientific},\ \bibinfo
  {year} {1988})\BibitemShut {NoStop}%
\bibitem [{\citenamefont {Song}\ \emph {et~al.}(2011)\citenamefont {Song},
  \citenamefont {Lazauskas},\ and\ \citenamefont {Gudkov}}]{Song:2010sz}%
  \BibitemOpen
  \bibfield  {author} {\bibinfo {author} {\bibfnamefont {Y.-H.}\ \bibnamefont
  {Song}}, \bibinfo {author} {\bibfnamefont {R.}~\bibnamefont {Lazauskas}}, \
  and\ \bibinfo {author} {\bibfnamefont {V.}~\bibnamefont {Gudkov}},\
  }\href@noop {} {\bibfield  {journal} {\bibinfo  {journal} {Phys. Rev.},\
  }\textbf {\bibinfo {volume} {C83}},\ \bibinfo {pages} {015501} (\bibinfo
  {year} {2011})}\BibitemShut {NoStop}%
\bibitem [{\citenamefont {Herczeg}(1966)}]{Pherzeg66}%
  \BibitemOpen
  \bibfield  {author} {\bibinfo {author} {\bibfnamefont {P.}~\bibnamefont
  {Herczeg}},\ }\href@noop {} {\bibfield  {journal} {\bibinfo  {journal} {Nucl.
  Phys.},\ }\textbf {\bibinfo {volume} {75}},\ \bibinfo {pages} {655} (\bibinfo
  {year} {1966})}\BibitemShut {NoStop}%
\bibitem [{\citenamefont {Herczeg}()}]{Herczeg:1987gp}%
  \BibitemOpen
  \bibfield  {author} {\bibinfo {author} {\bibfnamefont {P.}~\bibnamefont
  {Herczeg}},\ }\href@noop {} {}\bibinfo {note} {In Tests of Time Reversal
  Invariance in Neutron Physics, edited by N. R. Roberson, C. R. Gould and J.
  D. Bowman (World Scientific, Singapore, 1987), p.24.}\BibitemShut {Stop}%
\bibitem [{\citenamefont {Liu}\ and\ \citenamefont
  {Timmermans}(2004)}]{Liu:2004tq}%
  \BibitemOpen
  \bibfield  {author} {\bibinfo {author} {\bibfnamefont {C.~P.}\ \bibnamefont
  {Liu}}\ and\ \bibinfo {author} {\bibfnamefont {R.~G.~E.}\ \bibnamefont
  {Timmermans}},\ }\href@noop {} {\bibfield  {journal} {\bibinfo  {journal}
  {Phys. Rev.},\ }\textbf {\bibinfo {volume} {C70}},\ \bibinfo {pages} {055501}
  (\bibinfo {year} {2004})}\BibitemShut {NoStop}%
\bibitem [{\citenamefont {Faddeev}(1961)}]{Faddeev:1960su}%
  \BibitemOpen
  \bibfield  {author} {\bibinfo {author} {\bibfnamefont {L.~D.}\ \bibnamefont
  {Faddeev}},\ }\href@noop {} {\bibfield  {journal} {\bibinfo  {journal} {Sov.
  Phys. JETP},\ }\textbf {\bibinfo {volume} {12}},\ \bibinfo {pages} {1014}
  (\bibinfo {year} {1961})}\BibitemShut {NoStop}%
\bibitem [{\citenamefont {Lazauskas}\ and\ \citenamefont
  {Carbonell}(2004)}]{Lazauskas:2004hq}%
  \BibitemOpen
  \bibfield  {author} {\bibinfo {author} {\bibfnamefont {R.}~\bibnamefont
  {Lazauskas}}\ and\ \bibinfo {author} {\bibfnamefont {J.}~\bibnamefont
  {Carbonell}},\ }\href@noop {} {\bibfield  {journal} {\bibinfo  {journal}
  {Phys. Rev.},\ }\textbf {\bibinfo {volume} {C70}},\ \bibinfo {pages} {044002}
  (\bibinfo {year} {2004})}\BibitemShut {NoStop}%
\bibitem [{\citenamefont {Song}\ \emph {et~al.}(2007)\citenamefont {Song},
  \citenamefont {Lazauskas}, \citenamefont {Park},\ and\ \citenamefont
  {Min}}]{Song:2007bj}%
  \BibitemOpen
  \bibfield  {author} {\bibinfo {author} {\bibfnamefont {Y.-H.}\ \bibnamefont
  {Song}}, \bibinfo {author} {\bibfnamefont {R.}~\bibnamefont {Lazauskas}},
  \bibinfo {author} {\bibfnamefont {T.-S.}\ \bibnamefont {Park}}, \ and\
  \bibinfo {author} {\bibfnamefont {D.-P.}\ \bibnamefont {Min}},\ }\Doi
  {10.1016/j.physletb.2007.09.038} {\bibfield  {journal} {\bibinfo  {journal}
  {Phys. Lett.},\ }\textbf {\bibinfo {volume} {B656}},\ \bibinfo {pages} {174}
  (\bibinfo {year} {2007})}\BibitemShut {NoStop}%
\bibitem [{\citenamefont {Song}\ \emph {et~al.}(2009)\citenamefont {Song},
  \citenamefont {Lazauskas},\ and\ \citenamefont {Park}}]{Song:2008zf}%
  \BibitemOpen
  \bibfield  {author} {\bibinfo {author} {\bibfnamefont {Y.-H.}\ \bibnamefont
  {Song}}, \bibinfo {author} {\bibfnamefont {R.}~\bibnamefont {Lazauskas}}, \
  and\ \bibinfo {author} {\bibfnamefont {T.-S.}\ \bibnamefont {Park}},\ }\Doi
  {10.1103/PhysRevC.79.064002} {\bibfield  {journal} {\bibinfo  {journal}
  {Phys. Rev.},\ }\textbf {\bibinfo {volume} {C79}},\ \bibinfo {pages} {064002}
  (\bibinfo {year} {2009})}\BibitemShut {NoStop}%
\bibitem [{\citenamefont {Lazauskas}\ \emph {et~al.}(2009)\citenamefont
  {Lazauskas}, \citenamefont {Song},\ and\ \citenamefont
  {Park}}]{Lazauskas:2009nw}%
  \BibitemOpen
  \bibfield  {author} {\bibinfo {author} {\bibfnamefont {R.}~\bibnamefont
  {Lazauskas}}, \bibinfo {author} {\bibfnamefont {Y.-H.}\ \bibnamefont {Song}},
  \ and\ \bibinfo {author} {\bibfnamefont {T.-S.}\ \bibnamefont {Park}},\
  }\href@noop {} { (\bibinfo {year} {2009})},\ \Eprint
  {http://arxiv.org/abs/0905.3119} {arXiv:0905.3119 [nucl-th]} \BibitemShut
  {NoStop}%
\bibitem [{\citenamefont {Park}\ \emph {et~al.}(2003)\citenamefont {Park} \emph
  {et~al.}}]{Park:2002yp}%
  \BibitemOpen
  \bibfield  {author} {\bibinfo {author} {\bibfnamefont {T.~S.}\ \bibnamefont
  {Park}} \emph {et~al.},\ }\Doi {10.1103/PhysRevC.67.055206} {\bibfield
  {journal} {\bibinfo  {journal} {Phys. Rev.},\ }\textbf {\bibinfo {volume}
  {C67}},\ \bibinfo {pages} {055206} (\bibinfo {year} {2003})}\BibitemShut
  {NoStop}%
\bibitem [{\citenamefont {Pastore}\ \emph {et~al.}(2009)\citenamefont
  {Pastore}, \citenamefont {Girlanda}, \citenamefont {Schiavilla},
  \citenamefont {Viviani},\ and\ \citenamefont {Wiringa}}]{Pastore:2009is}%
  \BibitemOpen
  \bibfield  {author} {\bibinfo {author} {\bibfnamefont {S.}~\bibnamefont
  {Pastore}}, \bibinfo {author} {\bibfnamefont {L.}~\bibnamefont {Girlanda}},
  \bibinfo {author} {\bibfnamefont {R.}~\bibnamefont {Schiavilla}}, \bibinfo
  {author} {\bibfnamefont {M.}~\bibnamefont {Viviani}}, \ and\ \bibinfo
  {author} {\bibfnamefont {R.~B.}\ \bibnamefont {Wiringa}},\ }\Doi
  {10.1103/PhysRevC.80.034004} {\bibfield  {journal} {\bibinfo  {journal}
  {Phys. Rev.},\ }\textbf {\bibinfo {volume} {C80}},\ \bibinfo {pages} {034004}
  (\bibinfo {year} {2009})}\BibitemShut {NoStop}%
\bibitem [{\citenamefont {Girlanda}\ \emph {et~al.}(2010)\citenamefont
  {Girlanda} \emph {et~al.}}]{Girlanda:2010vm}%
  \BibitemOpen
  \bibfield  {author} {\bibinfo {author} {\bibfnamefont {L.}~\bibnamefont
  {Girlanda}} \emph {et~al.},\ }\href@noop {} {\bibfield  {journal} {\bibinfo
  {journal} {Phys. Rev. Lett.},\ }\textbf {\bibinfo {volume} {105}},\ \bibinfo
  {pages} {232502} (\bibinfo {year} {2010})}\BibitemShut {NoStop}%
\bibitem [{\citenamefont {Schiavilla}\ \emph {et~al.}(2008)\citenamefont
  {Schiavilla}, \citenamefont {Viviani}, \citenamefont {Girlanda},
  \citenamefont {Kievsky},\ and\ \citenamefont {Marcucci}}]{Schiavilla:2008ic}%
  \BibitemOpen
  \bibfield  {author} {\bibinfo {author} {\bibfnamefont {R.}~\bibnamefont
  {Schiavilla}}, \bibinfo {author} {\bibfnamefont {M.}~\bibnamefont {Viviani}},
  \bibinfo {author} {\bibfnamefont {L.}~\bibnamefont {Girlanda}}, \bibinfo
  {author} {\bibfnamefont {A.}~\bibnamefont {Kievsky}}, \ and\ \bibinfo
  {author} {\bibfnamefont {L.~E.}\ \bibnamefont {Marcucci}},\ }\href@noop {}
  {\bibfield  {journal} {\bibinfo  {journal} {Phys. Rev.},\ }\textbf {\bibinfo
  {volume} {C78}},\ \bibinfo {pages} {014002} (\bibinfo {year}
  {2008})}\BibitemShut {NoStop}%
\bibitem [{\citenamefont {Gudkov}()}]{Gudkov:1995tp}%
  \BibitemOpen
  \bibfield  {author} {\bibinfo {author} {\bibfnamefont {V.~P.}\ \bibnamefont
  {Gudkov}},\ }\href@noop {} {}\bibinfo {note} {In Parity and time reversal
  violation in compound nuclear states and related topics, edited by N.
  Auerbach and J. D. Bowman (World Scientific, Singapore, 1995),
  p.231.}\BibitemShut {Stop}%
\bibitem [{\citenamefont {Gudkov}(1992){\natexlab{b}}}]{Gudkov:1991tp}%
  \BibitemOpen
  \bibfield  {author} {\bibinfo {author} {\bibfnamefont {V.~P.}\ \bibnamefont
  {Gudkov}},\ }\href@noop {} {\bibfield  {journal} {\bibinfo  {journal}
  {Z.Phys.},\ }\textbf {\bibinfo {volume} {A343}},\ \bibinfo {pages} {437}
  (\bibinfo {year} {1992}{\natexlab{b}})}\BibitemShut {NoStop}%
\bibitem [{\citenamefont {Gudkov}\ \emph {et~al.}(1993)\citenamefont {Gudkov},
  \citenamefont {He},\ and\ \citenamefont {McKellar}}]{Gudkov:1992yc}%
  \BibitemOpen
  \bibfield  {author} {\bibinfo {author} {\bibfnamefont {V.~P.}\ \bibnamefont
  {Gudkov}}, \bibinfo {author} {\bibfnamefont {X.-G.}\ \bibnamefont {He}}, \
  and\ \bibinfo {author} {\bibfnamefont {B.~H.}\ \bibnamefont {McKellar}},\
  }\href@noop {} {\bibfield  {journal} {\bibinfo  {journal} {Phys.Rev.},\
  }\textbf {\bibinfo {volume} {C47}},\ \bibinfo {pages} {2365} (\bibinfo {year}
  {1993})}\BibitemShut {NoStop}%
\bibitem [{\citenamefont {Pospelov}(2002)}]{Pospelov:2001ys}%
  \BibitemOpen
  \bibfield  {author} {\bibinfo {author} {\bibfnamefont {M.}~\bibnamefont
  {Pospelov}},\ }\href@noop {} {\bibfield  {journal} {\bibinfo  {journal}
  {Phys. Lett.},\ }\textbf {\bibinfo {volume} {B530}},\ \bibinfo {pages} {123}
  (\bibinfo {year} {2002})}\BibitemShut {NoStop}%
\bibitem [{\citenamefont {Gudkov}(1990)}]{Gudkov:1990tb}%
  \BibitemOpen
  \bibfield  {author} {\bibinfo {author} {\bibfnamefont {V.~P.}\ \bibnamefont
  {Gudkov}},\ }\href@noop {} {\bibfield  {journal} {\bibinfo  {journal}
  {Phys.Lett.},\ }\textbf {\bibinfo {volume} {B243}},\ \bibinfo {pages} {319}
  (\bibinfo {year} {1990})}\BibitemShut {NoStop}%
\bibitem [{\citenamefont {Pospelov}\ and\ \citenamefont
  {Ritz}(2005)}]{Pospelov:2005pr}%
  \BibitemOpen
  \bibfield  {author} {\bibinfo {author} {\bibfnamefont {M.}~\bibnamefont
  {Pospelov}}\ and\ \bibinfo {author} {\bibfnamefont {A.}~\bibnamefont
  {Ritz}},\ }\href@noop {} {\bibfield  {journal} {\bibinfo  {journal} {Annals
  Phys.},\ }\textbf {\bibinfo {volume} {318}},\ \bibinfo {pages} {119}
  (\bibinfo {year} {2005})}\BibitemShut {NoStop}%
\bibitem [{\citenamefont {Baker}\ \emph {et~al.}(2006)\citenamefont {Baker}
  \emph {et~al.}}]{Baker:2006ts}%
  \BibitemOpen
  \bibfield  {author} {\bibinfo {author} {\bibfnamefont {C.~A.}\ \bibnamefont
  {Baker}} \emph {et~al.},\ }\href@noop {} {\bibfield  {journal} {\bibinfo
  {journal} {Phys. Rev. Lett.},\ }\textbf {\bibinfo {volume} {97}},\ \bibinfo
  {pages} {131801} (\bibinfo {year} {2006})}\BibitemShut {NoStop}%
\bibitem [{\citenamefont {Romalis}\ \emph {et~al.}(2001)\citenamefont
  {Romalis}, \citenamefont {Griffith},\ and\ \citenamefont
  {Fortson}}]{Romalis:2000mg}%
  \BibitemOpen
  \bibfield  {author} {\bibinfo {author} {\bibfnamefont {M.~V.}\ \bibnamefont
  {Romalis}}, \bibinfo {author} {\bibfnamefont {W.~C.}\ \bibnamefont
  {Griffith}}, \ and\ \bibinfo {author} {\bibfnamefont {E.~N.}\ \bibnamefont
  {Fortson}},\ }\href@noop {} {\bibfield  {journal} {\bibinfo  {journal} {Phys.
  Rev. Lett.},\ }\textbf {\bibinfo {volume} {86}},\ \bibinfo {pages} {2505}
  (\bibinfo {year} {2001})}\BibitemShut {NoStop}%
\bibitem [{\citenamefont {Dmitriev}\ and\ \citenamefont
  {Khriplovich}(2004)}]{Dmitriev:2003hs}%
  \BibitemOpen
  \bibfield  {author} {\bibinfo {author} {\bibfnamefont {V.}~\bibnamefont
  {Dmitriev}}\ and\ \bibinfo {author} {\bibfnamefont {I.}~\bibnamefont
  {Khriplovich}},\ }\href@noop {} {\bibfield  {journal} {\bibinfo  {journal}
  {Phys.Rept.},\ }\textbf {\bibinfo {volume} {391}},\ \bibinfo {pages} {243}
  (\bibinfo {year} {2004})}\BibitemShut {NoStop}%
\end{thebibliography}%

\end{document}